\documentclass{iopart}
\pdfoutput=1
\usepackage{graphicx}
\usepackage{amsfonts}
\usepackage{amssymb}
\usepackage{amsbsy}
\usepackage{latexsym}
\usepackage{color}
\usepackage{bm}

\newcommand{\text}[1]{\mathrm{#1}}
\newcommand{\ppp}{{}}
\newcommand{\pmp}{(+)}

\newcommand{\none}{\nonumber \\}
\newcommand{\half}{\frac{1}{2}}

\def\be {\begin{equation}}
\def\ee  {\end{equation}}
\def\bea {\begin{eqnarray}}
\def\eea {\end{eqnarray}}

\begin{document}

\title[An instability of hyperbolic space under the Yang-Mills flow]{An instability of hyperbolic space under \\ the Yang-Mills flow}

\author{Jack Gegenberg, Andrew C.\ Day, Haitao Liu  \\ and Sanjeev S.\ Seahra}

\address{Department of Mathematics \& Statistics, University of
New Brunswick \\ Fredericton, New Brunswick, E3B 5A3, Canada}

\begin{abstract} %
We consider the Yang-Mills flow on hyperbolic 3-space.  The gauge connection is constructed from the frame-field and (not necessarily compatible) spin connection components.  The fixed points of this flow include zero Yang-Mills curvature configurations, for which the spin connection has zero torsion and the associated Riemannian geometry is one of constant curvature.  Perturbations to the fixed point corresponding to hyperbolic 3-space can be expressed as a linear superposition of distinct modes, some of which are exponentially growing along the flow.  The growing modes imply the divergence of the (gauge invariant) perturbative torsion for a wide class of initial data, indicating an instability of the background geometry that we confirm with numeric simulations in the partially compactified case.  There are stable modes with zero torsion, but all the unstable modes are torsion-full. This leads us to speculate that the instability is induced by the torsion degrees of freedom present in the Yang-Mills flow.
\end{abstract}

\section{Introduction}

It has been known for a long time that a 3-dimensional theory of gravity can be constructed as a gauge theory from an action principle consisting of a \emph{Chern-Simons} (CS) functional for an appropriate gauge group \cite{carlip1}.  The theory is at least classically quite similar to Einstein gravity with an appropriate cosmological constant. Like general relativity in three dimensions, the CS gauge theory in vacuum has no local dynamical degrees of freedom.  It is very problematic to couple local matter degrees of freedom to a CS gauge theory---though topological matter can be so coupled \cite{cargeg}.

The Ricci flow \cite{rfrev} of a Riemannian metric $g_{ij}$ is $\dot g_{ij}=-2 R_{ij}(g)$, where $R_{ij}(g)$ is the Ricci curvature tensor of $g_{ij}$.  These partial differential equations are weakly parabolic, but can be made parabolic by adding an appropriate ``DeTurck term'' to the Ricci flow above.  The fixed points of this flow are precisely the Einstein spaces of the metric.  The parabolic character of these equations is what makes them such a powerful tool in geometric analysis.  For example, in three dimensions, they can be used to prove the Thurston/Poincare Uniformization Conjecture.  As well, they have been used to discuss the stability of and critical phenomena in  various physically and mathematically interesting geometries \cite{suneeta, Garfinkle:2003an,throat,Balehowsky:2010xk}.  In \cite{suneeta}, it was found that hyperbolic space (in any dimension) was linearly stable under Ricci flow.

In the above studies, the geometry was Riemannian---that is, the metric or frame-field was compatible with the spin connection---throughout the flow.  However, there is no {\it a priori} reason why the allowed geometries must be Riemannian.  In particular, one may wish to explore the possibility that the flow of the torsion could destabilize the geometry.

It is the goal of this paper to demonstrate that, at least in lower dimensions, the Yang-Mills flow, with appropriate gauge group, is a useful tool for examining a geometric flow which includes, in addition to the metric, the torsion of a linear connection built from that metric.  In Section 2, we review the construction of an appropriate gauge potential from the frame-field and not necessarily compatible spin connection.  In Section 3, we show that when such a gauge potential obeys the Yang-Mill (YM) flow, then the frame-field obeys a generalization of the Ricci flow, which includes the contribution from the torsion, while the spin connection flow is driven by the Cotton tensor with contributions from the Riemannian curvature.  If the torsion is turned off, then the frame-field flow reduces to the (normalized) Ricci flow.  In Section 4, the YM flow is linearized about fixed points with zero (gauge) curvature.  We conclude in Section 5 and discuss further work.

\section{The gauge potential}\label{sec:potential}

The construction of a gauge potential---that is, of an appropriate connection on a vector bundle over the spacetime manifold $M$---proceeds in three dimensions exactly as in Witten's construction \cite{at,witten} of a CS gauge theory.

We begin with a connection of the form:
\be
A:=A^{(+)b} T^+_b +A^{(-)b} T^-_b ,
\ee
where the $T^\pm_b$ are generators of the Lie group SU(2), with Lie algebra indices $a,b,..=1,2,3$.  The algebra of the $T^{\pm}_b$ is:
\be
[T^\pm_a,T^\pm_b]=\epsilon^c{}_{ab} T^\pm_c,  \quad [T^+_a,T^-_b]=0.
\ee
The $A^{(\pm) b} $ are su(2) valued 1-form fields over $M$ which transform under SU(2) gauge transformations as connections.

To connect with Riemannian geometry, we write
\be
A^{(\pm) b}= \omega^b \pm  C e^b,
\ee
where $C$ is a constant.  Now define new generators $J_b,P_b$ by
\be
J_b:=T^+_b + T^-_b, \quad P_b:=C(T^+_b - T^-_b).
\ee
We find that the algebra of the $J_b,P_b$ is:
\be
[J_a,J_b] = \epsilon^c{}_{ab} J_c, \quad
\left[J_a,P_b\right] = \epsilon^c{}_{ab} P_c, \quad
\left[P_a,P_b\right] = C^{2} \epsilon^c{}_{ab} J_c.
\ee

The curvature  of this connection is $F=F^{(+)} + F^{(-)}$, where
\be
F^{(\pm)}=dA^{(\pm)}+\frac{1}{2}\left[A^{(\pm)},A^{(\pm)}\right].
\ee
We find that
\be
F=\left( \Omega^a+\frac{C^2}{2}\epsilon^a{}_{bc} e^b\wedge e^c \right)J_a +\tau^a P_a,
\ee
where
\bea
\Omega^a := d\omega^a+\half\epsilon^a{}_{bc} \omega^b\wedge\omega^c, \\
\tau^a := de^a+\epsilon^a{}_{bc} \omega^b\wedge e^c.
\eea
We can see that it makes sense to interpret the $e^a$ as a frame field and $\omega^a$ as a spin connection (not in general compatible with the frame field) for a geometry.  Then $\Omega^a$ is the curvature and $\tau^a$ the corresponding torsion.  If the connection $A$ is flat (i.e. $F=0$), then the {\it geometry} is Riemannian (torsion free) and has constant curvature, negative, positive, zero, as $C$ is imaginary, real, zero, respectively.

Finally, we note that it is possible to express the torsion directly in terms of the gauge potential as follows:
\begin{equation} \fl
	\tau^{a} = \frac{ \left( dA^{(+)a} + \frac{1}{2} \epsilon^{a}{}_{bc} A^{(+)b} \wedge A^{(+)c} \right) - \left( dA^{(-)a} + \frac{1}{2} \epsilon^{a}{}_{bc} A^{(-)b} \wedge A^{(-)c} \right)}{2C}.\label{eq:torsion}
\end{equation}
We will use this when we consider perturbations about globally hyperbolic space.

\section{The Yang-Mills flow}

We consider a one-parameter family of connections given locally by the gauge potential $A(t,x^{1},x^{2},x^{3})$, where $t$ is the flow parameter, and the $x^i$ are coordinates on a smooth 3-dimensional manifold $M$.  The Yang-Mills-DeTurck  flow \cite{ab,don,rade,bg} is given by \footnote{The first use of the YM flow with the gauge potential constructed from the frame-field and spin connection was in \cite{bg}. }
\be \label{eq:full flow}
\dot{A}=-\star D \star F - Dv,
\ee
where $\star$ is the Hodge dual of the form field, $\dot A:=\partial_t A$, and the 0-form field $v:=v^{(+) b}T^{+}_b+v^{(-) b}T^{-}_b$ is the DeTurck field.\footnote{ Note that the presence of the arbitrary DeTurck field $v$ in the flow equation (\ref{eq:full flow}) reflects the dependence of the potential $A$ on the choice of gauge.
}  To define the Hodge dual, one needs a metric.  For a 2-form $F:=\half F_{ij} dx^i\wedge dx^j$ in three dimensions,
\be
\star F=\half F_{ij}\bar{\eta}^{ij}{}_k dx^k,
\ee
where $\bar{\eta}^{ijk}$ is the Levi-Civita {\it tensor} defined with respect to a background Riemannian metric $\bar{g}_{ij}$:
\be
\bar{\eta}^{ijk}=\frac{\epsilon^{ijk}}{\sqrt{\bar g}}.
\ee
Above, $\bar g$ is the determinant of $\bar{g}_{ij}$ and $\epsilon^{ijk}$ is the permutation symbol defined so that $\epsilon^{123}=+1$. Coordinate (that is, tangent plane) indices are raised and lowered by the metric tensor $\bar{g}_{ij}$.
The gauge covariant derivative $D$ is defined to be
\be
D\sigma:=d\sigma +\left[A, \sigma\right],
\ee
for $\sigma$ a Lie algebra valued $p$-form field, i.e. $\sigma=\sigma^{(+)a}T_a^+ + \sigma^{(-) a}T^-_a$.

Given the independence of the generators of the Lie group, the Yang-Mills flow splits naturally into two flows, written in terms of $e^a$ and $\omega^a$ as follows:
\bea \fl
\dot\omega^a = -\star\left[ C^2(\tau^a+\epsilon^a{}_{bc}e^b\wedge\star\tau^c) +D_\omega \star\Omega\right]
 - D_\omega \beta^a-C^2\epsilon^a{}_{bc}e^b\alpha^c; \\ \fl
\dot e^a = \star\left[ \epsilon^a{}_{bc}e^b\wedge(\star\Omega^c+C^2 e^c) -D_\omega \star\tau^a\right]-D_\omega \alpha^a-\epsilon^a{}_{bc}e^b\beta^c.
\eea
These equations are, respectively, the coefficients of $J_a$ and $P_a$ in the YM equations.  We have defined
\be
D_\omega\sigma^a :=d \sigma^a+\epsilon^a{}_{bc}\omega^b \wedge\sigma^c.
\ee
The Lie algebra valued 0-forms $\alpha^a,\beta^a$ are constructed from the DeTurck 0-form field as follows:
\bea
\alpha^a:&=&\half\left(v^{(+)a}-\frac{1}{C}v^{(-)a}\right);\none
\beta^a:&=&\frac{C}{2}\left(v^{(+)a}+\frac{1}{C}v^{(-)a}\right),
\eea
where $v:=v^{(+)a}T^{+}_a+v^{(-)a}T^{-}_a$.

We now examine a special class of fixed points of this flow which have:
\begin{enumerate}
\item vanishing DeTurck field $v^{a} = 0$;
\item vanishing Yang-Mills curvature $F=0$; and \label{cond:2}
\item metric $\bar{g}_{ij}=e^a_i e^b_j \delta_{ab}$ with $\det{e^a_i}\neq 0$; i.e., a metric derived from the frame fields. \label{cond:3}
\end{enumerate}
Conditions (\ref{cond:2}) and (\ref{cond:3}) imply that such fixed points have vanishing torsion $\tau^{a} = 0$ and constant Riemannian curvature.  In this case we find that the right hand sides of the flow equations reduce to
\bea \label{eq:background field eqn}
R_{ij}(\bar{g})+2 C^2 \bar {g}_{ij}=0;\\
\eta^{ij}{}_k \nabla_i G_{jl}(\bar {g})=0.
\eea
Here, $R_{ij}(\bar{g})$ and $G_{ij}(\bar{g})$ are the Ricci and Einstein tensors of the background metric $\bar{g}_{ij}$, respectively.  The second equation expresses the vanishing of the Cotton tensor.  Note that by virtue of the first equations (the Einstein equations), the Cotton tensor identically vanishes.

Thus, with the above choice of the background metric, the Einstein manifolds are fixed points.  This sets up the problem of examining the linear stability of those fixed points, which include the sphere/hyperbolic space/flat space (depending on the gauge group).  Furthermore, for the case of negative curvature, the (Euclideanized) black hole geometries are also fixed points.  Of course there are other fixed points which are not torsion-free geometries.  Our task here is to begin the process of mapping the configuration space of flowing geometries, first of all in the vicinity of the fixed points which are Einstein spaces.

\section{Linearized flow}

\subsection{Perturbative equations of motion}

In this section, we consider the Yang-Mills flow of $A$ in the vicinity of torsion-free fixed points with $F=0$.  We note that since $[T^+, T^-]=0$, the flow equations governing $A^{(+)}$ and $A^{(-)}$ are decoupled:
\be
\dot A^{(\pm)}=-\star D^{(\pm)}\star F^{(\pm)}-D^{(\pm)} v^{(\pm)},
\ee
where the action of the gauge covariant derivative on an arbitrary $p$-form is  $D^{(\pm)}\sigma=d\sigma +[A^{(\pm)},\sigma]$.  To ease readability, we will temporarily suppress the $(\pm)$ superscript on all quantities throughout most of this subsection.  Hence, all the following formulae apply equally to the ``$+$'' or ``$-$'' sectors.

We perturb the fixed-point gauge potential as follows:
\begin{equation}
	A^{b}_{i} \rightarrow A^{b}_{i} + \delta A^{b}_{i}, \quad \delta A^{b}_{i} = u^{b}_{i}.
\end{equation}
In general, $u^{\ppp b}=u^{\ppp b}_i(t,x^{j})dx^i$ will perturb both the background frame field and the spin connection, making the latter torsion-full.  To linear order in the perturbations, the flow equations reduce to
\be
\dot {u}^{\ppp a}_i=D^{\ppp j} D^{\ppp}_j u^{\ppp a}_i-D^{\ppp j} D^{\ppp}_i u^{\ppp a}_j - D_{i} v^{a},\label{dotu1}
\ee
where the gauge covariant derivative is
\be
D_i u^{\ppp a}_j:=\nabla_i u^{\ppp a}_j +\epsilon^a{}_{bc}A^{\ppp b}_i u^{\ppp c}_j.
\ee
We have assumed that the DeTurck field vanishes in the unperturbed solution, and hence $v^{a}$ is the same order as $u^{a}_{i}$.   Notice that the perturbation to the field strength is $\delta F^{\ppp a}_{ij} =  2D^{\ppp}_{[i}u^{\ppp a}_{j]}$, so the perturbative flow equations can be written as $\dot{u}^{\ppp a}_{i} = - D^{\ppp j} \delta F^{\ppp a}_{ij} - D_{i}v^{a}$.  In the above, $\nabla_i$ is the covariant derivative with respect to the background Riemannian metric $\bar{g}_{ij}$.  Since all the partial derivatives $\partial_i$ of 1-forms are skew-symmetrized, we can replace them by the covariant derivatives $\nabla_i$ with respect to the background metric.

By commuting derivatives in the second term on the RHS of (\ref{dotu1}), we can re-write the flow equation as
\begin{equation}
\dot{u}^{\ppp a}_i=D^{\ppp j} D^{\ppp}_j u^{\ppp a}_i -R_i{}^{j} u^{\ppp a}_j - D_{i}(v^{a}+L^{a}),
\end{equation}
where we have defined $L_{a} = D^{i} u_{i}^{a}$.  If we now choose the DeTurck field such that:
\begin{equation}\label{eq:DeTurck}
	D_{i} v^{a} = -D_{i} L^{a},
\end{equation}
the resulting flow equation resembles a Yang-Mills version of the heat equation
\begin{equation}
\dot{u}^{\ppp a}_i=D^{\ppp j} D^{\ppp}_j u^{\ppp a}_i -R_i{}^{j} u^{\ppp a}_j, \label{dotu2}
\end{equation}
with ``mass'' term $R_i{}^{j} u^{\ppp a}_j = -2C^{2} u^{\ppp a}_i$ (assuming we perturb about an $F=0$ solution).

It is important to note that the specification of the DeTurck field (\ref{eq:DeTurck}) does not necessarily remove all gauge arbitrariness from the solutions of (\ref{dotu2}).  Let us consider an infinitesimal gauge transformation generated by the Lie-algebra valued scalar $\psi^{a}$.  Under such a transformation, the gauge potential perturbation transforms as
\begin{equation}
    u^{\ppp a}_{i} \mapsto u^{\ppp a}_{i} + D^{\ppp}_{i}\psi^{\ppp a} = u^{\ppp a}_{i} + \nabla_{i}\psi^{\ppp a} + \epsilon^{a}{}_{bc}A^{\ppp b}_{i} \psi^{\ppp c},
\end{equation}
After some algebra, we find that the flow equation (\ref{dotu2}) transforms as
\begin{equation}
\dot{u}^{\ppp a}_i=D^{\ppp j} D^{\ppp}_j u^{\ppp a}_i -R_i{}^{j} u^{\ppp a}_j -D_{i}(\dot\psi^{a} - D_{j} D^{j} \psi^{a}).
\end{equation}
Hence, the flow equation (\ref{dotu2}) is invariant under a restricted class of gauge transformations with generators satisfying
\begin{equation}
	D_{i}(\dot\psi^{a} - D_{j} D^{j} \psi^{a}) = 0.
\end{equation}
Whether or not there exists any non-trivial $\psi^{a}$ satisfying this equation depends on the nature of the background geometry and boundary conditions.

We note that under a gauge transformation, the perturbation to the field strength $\delta F^{\ppp a}_{ij} =  2D^{\ppp}_{[i}u^{\ppp a}_{j]}$ transforms as
\begin{equation}
    \delta F^{\ppp a}_{ij}  \mapsto \delta F^{\ppp a}_{ij} + \underbrace{\epsilon^{a}{}_{bc} \psi^{\ppp b} F_{ij}^{\ppp c}}_{=0},
\end{equation}
where we have used that $F=0$ in the background.  Hence, $\delta F^{\ppp a}_{ij}$ is gauge invariant.  It can be shown that $\delta F^{a}_{ij}$ obeys its own flow equation
\begin{equation}\label{eq:F EOM}
	\dot{\delta F^{\ppp a}_{ij}} = (D^{k}D_{k} + 2C^{2}) \delta F^{\ppp a}_{ij}.
\end{equation}
Finally, we examine the linear perturbations to the torsion tensor (\ref{eq:torsion}).  Restoring the ``$+$'' and ``$-$'' superscripts, we find
\begin{equation}\label{eq:perturbative torsion}
	\delta\tau^{a}_{ij} = \frac{1}{C} \left\{ D^{(+)}_{[i} u^{(+)a}_{j]} - D^{(-)}_{[i} u^{(-)a}_{j]} \right\}. 	
\end{equation}
It follows that $\delta\tau^{a}_{ij}$ is gauge invariant if $F= 0$ in the background.  When combined (\ref{eq:F EOM}), this implies that if the initial data for a perturbation is torsion-free, the torsion will remain zero along the flow.

\subsection{Fluctuations about globally hyperbolic space}

We now apply the general formulae of the previous subsection to flows in the vicinity of 3-dimensional hyperbolic space.  The background metric is
\be
ds^2=\frac{dx^{2}+dy^{2}+dz^{2}}{k^2 y^2},
\ee
where $k$ is a constant which fixes the ``size'' of the manifold;  that is, it is the square root of the absolute value of the cosmological constant.  The background frame field and corresponding (torsion-free) spin connection are
\begin{equation}
	e^{a} = \frac{1}{ky} \left[ dx, dy, dz \right], \quad \omega^{a} = \frac{1}{y} \left[ -dz, 0, dx \right],
\end{equation}
respectively.  The background gauge potential is
\be
A^{(\pm) b}= \omega^b \pm  i k e^b;
\ee
i.e., the constant introduced in \S\ref{sec:potential} above is
\begin{equation}\label{eq:C}
C = ik.
\end{equation}
Since $\omega^{a}$, $e^{a}$ and $k$ are real, we find
\begin{equation}\label{eq:conjugate}
A^{(+)} = [A^{(-)}]^{*}, \quad u^{(+)a}_{i} = [u^{(-)a}_{i}]^{*},
\end{equation}
which means that the entire perturbation is specified by the nine (complex) components of $u^{(+)a}_{i}$.  Hence, we only consider the ``$+$'' part of the perturbation below.

It is useful to re-arrange the components of $u_{i}^{\pmp a}$ into a nine-dimensional vector as follows:
\begin{equation}
	\fl \vec{\xi} = [u_{2}^{\pmp 1},u_{1}^{\pmp 2},u_{3}^{\pmp 2},u_{2}^{\pmp 3},u_{1}^{\pmp 1},u_{3}^{\pmp 1},u_{2}^{\pmp 2},u_{1}^{\pmp 3},u_{3}^{\pmp 3}]^{T}.
\end{equation}
For simplicity, we assume that $u_{i}^{\pmp a}$ does not depend on $x$ or $z$; that is, $\vec{\xi} = \vec{\xi}(t,y)$. Then, if we change the spatial coordinate to
\begin{equation}
	w = \ln(ky),
\end{equation}
it is possible to write the most general solution of (\ref{dotu2}) that is independent of $x$ and $z$ in terms of Fourier modes:
\begin{eqnarray}\none
	\vec{\xi}(t,w) = \sum_{\alpha=1}^{9}\int_{-\infty}^{\infty} d\eta \, a_{\alpha}(\eta) \vec\xi_{\alpha}(t,w;\eta), \\ \vec\xi_{\alpha}(t,w;\eta) = \vec{\zeta}_{\alpha}(\eta)  \exp[ \lambda_{\alpha}(\eta)k^{2} t + i\eta w],\label{eq:pert soln}
\end{eqnarray}
where the $a_{\alpha}(\eta)$ are nine complex Fourier amplitudes.  Here, $\lambda_{\alpha}$ and $\vec{\zeta}_{\alpha}$ are solutions of the eigenvalue problem ${\cal A} \vec{\zeta}_{\alpha} = \lambda_{\alpha} \vec{\zeta}_{\alpha}$, where the block-diagonal matrix ${\cal A} = {\cal A}(\eta)$ is given by:
\begin{equation}
        {\cal A} =
	\left(\begin{array}{c|c}
	       {\cal A}_{1} & \\
	       \hline & {\cal A}_{2}
	\end{array}
	\right),
\end{equation}
with
\begin{equation}\fl
	{\cal A}_{1} =
	\left(\begin{array}{cccc}
	-\eta^{2} & -2 & -2i & -2\eta \\
	-2 & -\eta^{2} & 0 & 2i \\
	-2i & 0 & -\eta^{2} & -2 \\
	2\eta & 2i & -2 & -\eta^{2}
	\end{array}\right)
	\quad {\cal A}_{2} =
	\left(\begin{array}{ccccc}
	1-\eta^{2} & 0 & 2 & -2\eta & 0 \\
	0 & 1-\eta^{2} & 2i & 0 & -2\eta \\
	2 & 2i & -1-\eta^{2} & -2i & 2 \\
	2\eta & 0 & -2i & 1-\eta^{2} & 0 \\
	0 & 2\eta & 2 & 0 & 1-\eta^{2}
	\end{array}\right).
\end{equation}
The block diagonal structure of $\mathcal{A}$ implies that the components of $u_{i}^{(+)a}$ with $i+a$ even are decoupled from the components with $i+a$ odd.  We find that $\mathcal{A}$ possesses nine linearly independent eigenvectors, and we summarize the important features of the associated eigenmodes $\vec\xi_{\alpha}(t,w;\eta)$ in Table \ref{tab:eigenmodes}.
\begin{table*}
\begin{tabular*}{\textwidth}{c@{\extracolsep{\fill}}cccc}\hline
eigenmode & $\lambda_{i}$ & stability & torsion \\
\hline\hline$1,2$ & $-\eta^{2} \pm 2i\eta$ & stable & torsion-free \\
$3,4$ & $-\eta^{2}$ & stable & torsion-full \\
$5$ & $-1-\eta^{2}$ & stable & torsion-free \\
$6,7$ & $1-2i\eta -\eta^{2}$ & unstable for $|\eta| < 1$ & torsion-full \\
$8,9$ & $1+2i\eta -\eta^{2}$ & unstable for $|\eta| < 1$ & torsion-full \\ \hline
\end{tabular*}
\caption{Features of the eigenmode solutions $\vec\xi_{i}$ for perturbations to the gauge potential in the globally hyperbolic case.  It can be shown that the torsion-free modes ($i=1,2,5$) also have $D_{[i} u_{j]}^{a} = 0$ and can be written as the gauge-covariant gradient of Yang-Mills valued scalars: $u_{i}^{a} = D_{i} \Lambda^{a}$.  This suggests that there exists a local gauge transformation such that $u_{i}^{a} \mapsto 0$; i.e., these solutions are pure gauge.  Whether or not such a transformation is globally defined is an open question.}\label{tab:eigenmodes}
\end{table*}

From Table \ref{tab:eigenmodes}, we see that four of the modes $\{ \vec\xi_{6},\vec\xi_{7},\vec\xi_{8},\vec\xi_{9} \}$ have $\text{Re}(\lambda_{\alpha}) > 0$ for $|\eta| < 1$.   Since $\vec\xi_{\alpha} \propto e^{\lambda_{\alpha}k^{2}t}$, this implies that these modes are exponentially growing for long wavelengths, suggesting an instability.    To gain further insight, we can evaluate the torsion associated with each of the eigenmodes.  Using (\ref{eq:perturbative torsion}), (\ref{eq:C}) and (\ref{eq:conjugate}), we obtain
\begin{equation}
	\delta\tau^{a}_{ij} = 2k^{-1} \text{Im} \left\{ D^{(+)}_{[i} u^{(+)a}_{j]} \right\}. 	
\end{equation}
We find that the unstable modes $\{ \vec\xi_{6},\vec\xi_{7},\vec\xi_{8},\vec\xi_{9} \}$ all have nonzero torsion that is exponentially growing along the flow for $|\eta| < 1$.  Since $\tau_{ij}^{a}$ is gauge invariant, we can conclude that the suspected instability is not an artifact of the DeTurck gauge fixing (\ref{eq:DeTurck}).

Conventionally, a fixed point is said to be unstable under a geometric flow if we can find a solution with finite but growing norm \cite{suneeta}; or equivalently, a growing solution satisfying appropriate boundary conditions.  It is fairly easy to use the above eigenmodes to obtain solutions to (\ref{dotu2}) that satisfy Dirichlet boundary conditions as $w \rightarrow \pm\infty$.  For example, the $9^\text{th}$ eigenmode has:
\begin{equation}
	\vec\zeta_{9} = [0,0,0,0,0,0,k,0,0,k]^{T}, \quad \lambda_{9} = 1+2i\eta -\eta^{2}.
\end{equation}
Plugging this into the expansion (\ref{eq:pert soln}) and choosing $a_{\alpha} = \delta_{\alpha,9}$ (which corresponds to a particular choice of initial data) we generate the following solution to the flow equations:
\begin{equation}\label{eq:particular solution}
	u^{(+)a} = \sqrt{\frac{\pi}{k^{2} t}} \exp \left\{ -\frac{w(4k^{2}t+w)}{4k^{2}t} \right\}[dx,0,-i\,dx].
\end{equation}
For $t >0$, this solutions satisfies $u^{(+)a} \rightarrow 0$ for $w \rightarrow \pm\infty$ (or $y \rightarrow 0$ and $y \rightarrow \infty$).  We define the pointwise tensor norm of the perturbation as
\begin{equation}
	\Vert u^{(+)} \Vert^{2} = \int \text{Tr} \left[ u^{(+)*} \wedge \star \, u^{(+)} \right]  = \frac{k\mathcal{V}_{\!\perp}(2\pi)^{3/2} e^{2k^{2}t}}{(k^{2}t)^{1/2}},
\end{equation}
where
\begin{equation}
	\mathcal{V}_{\!\perp} = \int\!\!\!\!\int dx \, dz
\end{equation}
is the ``volume'' of the transverse dimensions.\footnote{One can define many other norms, \emph{cf}.\ \cite{suneeta} for several possibilities.}  Assuming that $\mathcal{V}_{\!\perp} < \infty$, we see that $\Vert u^{(+)} \Vert$ is finite for $t \in (0,\infty)$, but diverges as $t \rightarrow \infty$.  Hence, (\ref{eq:particular solution}) represents a normalizable and unstable solution of the flow equations.  Note that there is nothing particularly special about the choice $a_{\alpha} = \delta_{\alpha,9}$; many choices of the Fourier amplitudes---or, equivalently, initial data---will excite unstable but normalizable solutions.

%It is important to note that these unstable eigenmodes are to be understood in the context of a linear superposition of the form (\ref{eq:pert soln}).  Therefore, even though the eigenmodes are not themselves normalizable, they will in general contribute to a normalizable solution of the flow equations (provided that the Fourier amplitudes $a_{\alpha}(\eta)$ satisfy appropriate fall-off conditions).\footnote{This is highly analogous to the situation in quantum mechanics where one constructs normalizable wave packets from unnormalizable plane waves.}  In fact, a broad class of initial data for the perturbation with compact support will have non-zero overlap with the ``dangerous'' modes (i.e.\ have $a_{\alpha}(\eta) \not\equiv 0$ for $|\eta|<1$ and $\alpha=6,7,8,9$) and hence will excite the instability.  This will also hold for perturbations with non-trivial profiles in the $x$ and $z$ directions.  That is, we expect most solutions of (\ref{dotu2}) corresponding to normalizable initial data to be unstable.

\begin{figure}
\begin{center}
\includegraphics[width=0.6\textwidth]{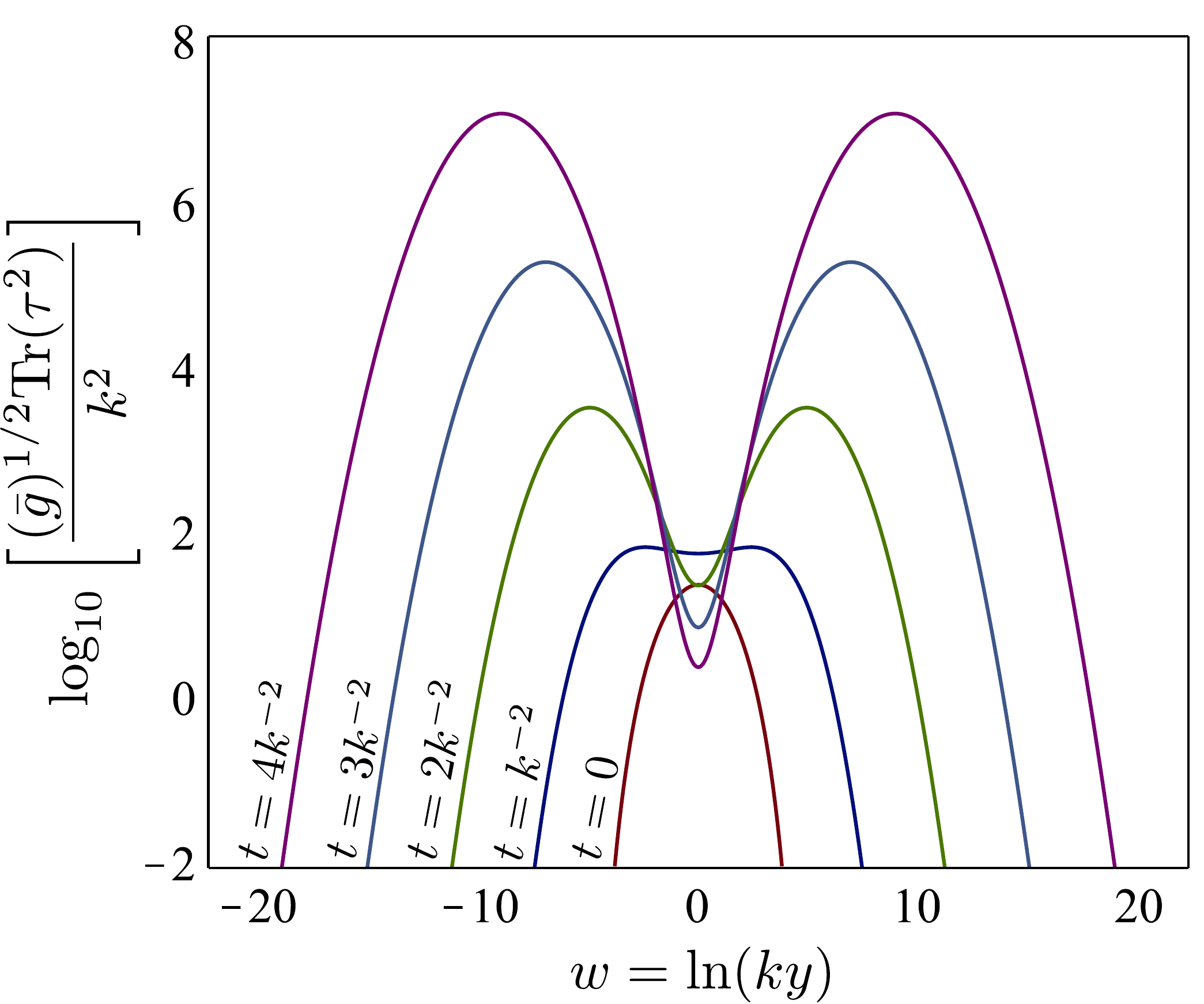}
\end{center}
\caption{An example of the evolution of the (densitized) trace of the torsion tensor squared under the Yang-Mills flow linearized about hyperbolic space.  The curves were obtained from a Crank-Nicholson numeric solution of (\ref{dotu2}), and we have assumed that the pertubation does not depend on $x$ or $z$.  Our choice of initial data (given by the $t=0$ curve) has compact support, and we see from the observed growth of the peaks that the torsion amplitude grows exponentially with the flow parameter $t$.  Since the torsion tensor is invariant, these profiles will look the same in any gauge.  We find that the instability is excited by a wide class of initial data.}\label{fig}
\end{figure}

To illustrate this point, we have numerically solved the flow equation (\ref{dotu2}) using a Crank-Nicholson finite difference scheme.  We take the perturbation to be independent of $x$ and $z$, and we assume compact support initial data for $u_{i}^{a}$ for $w \in(-\infty,\infty)$.  In figure \ref{fig}, we plot the flow evolution of the square of the torsion tensor $\text{Tr}( \tau^{2})$, which is gauge invariant.  The numeric solution is clearing growing exponentially along the flow, vividly demonstrating the instability.

If $x$ and $z$ are non-compact dimensions, the transverse volume $\mathcal{V}_{\!\perp}$ and hence the norm of solutions like (\ref{eq:particular solution}) will be infinite.  However, we still expect generic initial data with finite norm to have non-zero overlap with the ``dangerous'' eigenmodes listed in Table \ref{tab:eigenmodes} and hence excite the instability.  That is, it is likely that the fully non-compact hyperbolic 3-plane is unstable under the Yang-Mills flow, but this needs to be confirmed by explicit calculation.

\section{Discussion}

We constructed the Yang-Mills flow on a 3-manifold for the case that the gauge potential is decomposed into a frame-field and spin connection part.  The fixed points of this flow include the spaces of constant curvature, with the sign of the curvature determined by the gauge group under consideration.  We have shown analytically and numerically that partially compactified 3-dimensional hyperbolic space is unstable under this YM flow, and all linearly unstable solutions have nonzero torsion.  Furthermore, we expect this instability to persist in the fully non-compact case.  This should be compared to earlier discussions \cite{suneeta}, where it was found that under the Ricci flow (with zero torsion) hyperbolic space was linearly stable.  We are hence led to they conjecture that it is the torsion inherent in the Yang-Mills flow that is responsible for the instability.

We are currently extending our analysis to the question of the stability of 3D black hole geometries, and other fixed points of the YM flow constructed from ``geometrized'' gauge potentials.  It remains to be seen if there are any stable fixed points, or if the flow is singular.  The outcome will be relevant to the starting point of any attempt to construct quantum gravity theories in three and perhaps higher dimensions.

\ack We would like to thank G. Kunstatter and V.~Suneeta for valuable discussions. The authors wish to acknowledge the financial support of the Natural Sciences and Engineering Research Council (NSERC) of Canada.  H.\ Liu wishes to thank G. Kunstatter for his hospitality during his stay at the University of Winnipeg, where some of this work was done.

\bigskip

\providecommand{\href}[2]{#2}\begingroup\raggedright\endgroup

\end{document}